**Selective colloid transport across planar polymer brushes**

*Mikhail Y. Laktionov[1], Ekaterina B. Zhulina[1,2], Leonid Klushin[2,3], Ralf P. Richter[4],\*, Oleg V. Borisov[1,2,5],\**

This work is dedicated to the memory of Wolfgang Meier, an outstanding scientist and a wonderful person. His pioneering works on stimuli-responsive triblock copolymer polymersomes paved the way for design of artificial biomimetic polymeric nanostructures, largely motivating theoretical studies of relationships of their self-organization and interactions with bionanocolloids. The here-presented work can be considered as one further step in this exciting direction, opened up by Wolfgang Meier.

[1] M. Y. Laktionov, E. B. Zhulina, O. V. Borisov

Petersburg National Research University of Information Technologies, Mechanics and Optics, 197101 St. Petersburg, Russia

[2] E. B. Zhulina, L. Klushin, O. V. Borisov

Institute of Macromolecular Compounds, Russian Academy of Sciences, 31 Bolshoy pr, 199004 Saint Petersburg, Russia

[3] L. Klushin

Department of Physics, American University of Beirut, P.O. Box 11-0236, Beirut 1107 2020, Lebanon

[4] R. P. Richter

School of Biomedical Sciences, Faculty of Biological Sciences, School of Physics and Astronomy, Faculty of Engineering and Physical Sciences, Astbury Centre for Structural Molecular Biology, and Bragg Centre for Materials Research, University of Leeds, Leeds LS2 9JT, United Kingdom

\* Email: r.richter@leeds.ac.uk

[5] O. V. Borisov

CNRS, Université de Pau et des Pays de l'Adour UMR 5254, Institut des Sciences Analytiques et de Physico-Chimie pour l'Environnement et les Matériaux, 64053 Pau, France

* Email: oleg.borisov@univ-pau.fr

Polymer brushes are attractive as surface coatings for a wide range of applications, from fundamental research to everyday life, and also play important roles in biological systems. How colloids (*e.g.*, functional nanoparticles, proteins, viruses) bind and move across polymer brushes is an important yet under-studied problem. We present a mean-field theoretical approach to analyze the binding and transport of colloids in planar polymer brushes. The theory explicitly considers the effect of solvent strength on brush conformation and of colloid-polymer affinity on colloid binding and transport. We derive the position-dependent free energy of the colloid insertion into the polymer brush which controls the rate of colloid transport across the brush. We show how the properties of the brush can be adjusted for brushes to be highly selective, effectively serving as tuneable gates with respect to colloid size and affinity to the brush-forming polymer. The most important parameter regime simultaneously allowing for high brush permeability and selectivity corresponds to a condition when the repulsive and attractive contributions to the colloid insertion free energy nearly cancel. Our theory should be useful to design sensing and purification devices with enhanced selectivity and to better understand mechanisms underpinning the functions of biopolymer brushes.

# 1. INTRODUCTION

Solvated polymer films are attractive as surface coatings for a wide range of applications, from fundamental research to everyday life, and also play important functional roles in biological systems. One of the most popular implementations of such coatings are polymer brushes, that is, films of polymer chains (linear and flexible in their simplest form) that are grafted *via* one of their two ends to the surface at a density that is sufficiently to generate a 'forest' of dynamically moving chains that evenly covers the surface[1, 2].

A salient functional parameter of solvated polymer films at interfaces is their ability to selectively bind and transport macromolecular agents diffusing in the solvent phase. This is clearly important in biological systems. For example, the bulk transport of proteins and nucleic acids between the nucleus and the cytosol of eukaryotic cells, critical for orderly gene transcription and translation and thus basic cell function, is controlled by nanoscale channels (the nuclear pore complexes) that perforate the nuclear envelope and are filled with a brush of specialized intrinsically disordered protein chains[3]. Also, many cells are coated with a 'forest' of carbohydrates called the glycocalyx[4, 5], which modulates access of extracellular signaling molecules (*e.g.*, morphogens, cytokines and exosomes) that guide cell communication and behavior. The glycocalyx is also the first line of defense against exogenous agents (although many viruses and toxins have evolved to bind to glycocalyces to define their host cell specificity and facilitate cell entry[6, 7]) and plays important roles in filtration of waste products (*e.g.*, in glomerular filtration in the kidneys[8]), and in preventing ingress of toxic protein species (*e.g.*, of α-synuclein, the causative agent of Parkinson's disease, in neurons[9]). Similarly, a coating rich in lipopolysaccharides helps to protect gram-negative bacteria against antibiotics and antimicrobial peptides[10].

Selective binding and transport is also desirable for technological applications. In particular, polymer brushes can be envisaged as a selective barrier in biosensor applications[11], by preventing the access of undesired molecules and/or facilitating access of the desired agent, to the biosensor surface thus enhancing biosensor selectivity and/or sensitivity. Similarly, one can envisage solvated polymer films to coat the outer surface of filtration membranes to impart superior selectivity compared to conventional membranes[12], or to the surface of nanoscopic carriers (*e.g.*, liposomes and polymersomes) to control the sustained release of active agents for biomedical applications. For example, polymersomes self-assembled from triblock copolymers, with an insoluble central block that constitutes the core of the polymer membrane and two soluble terminal blocks that form brush-like coronae on either side, were synthesized and explored by Wolfgang Meier *et al.* [13-15]. Understanding the permeation properties of the inner and outer brush layers would benefit the application of polymersomes as carriers for controlled agent (*e.g.*, drug and gene) delivery, or as nanoreactors and artificial organelles[16, 17]. The chemical design space of polymers is vast, and with appropriate tuning of polymer functionality it should be possible to impart a level of selectivity on polymer brushes that goes well beyond separation by crude parameters such as the size or overall charge of the macromolecular agent.

The rational design of polymer brushes for this type of applications, however, remains challenging, and our understanding of the mechanisms underpinning agent transport in biological polymer coatings is also limited. For (i) a macromolecular agent (henceforward called colloid) of given size and surface chemistry, (ii) a brush of given polymer chemistry and grafting density, and (iii) a given solvent, it remains nontrivial to predict how strongly the agent will bind and how fast it will diffuse in the brush. Qualitatively, colloid binding to brushes can be understood as a process involving two opposing effects: entropic excluded volume (*i.e.*, osmotic) effects repel the colloid, whereas adhesion to polymers attracts the colloid. Both effects increase with colloid size and polymer density, albeit to different degrees, and it is the subtle balance of two large yet opposing effects that defines whether there is a net attraction or repulsion. Moreover, brushes are sensitively affected by the solvent strength. The effect of solvent strength on brush morphology (and in particular on the polymer density profile) is rather well understood[18, 19], but we know less about how the solvent strength impacts on colloid binding. Lastly, the diffusive transport of colloids across polymer brushes has received relatively little attention so far, and open questions are how the solvent strength and colloid-polymer interactions jointly define colloid transport across the brush, and how colloid binding and transport can be made selective.

Here, we present a theoretical mean-field approach to analyze the binding and transport of colloids in polymer brushes. For clarity, we focus on the specific case of planar polymer brushes; the main predictions, however, are also applicable to other brush geometries. In seeking to identify the basic parameters required for selective binding and transport, we explicitly consider the effect of solvent strength on brush conformation (and notably polymer volume fraction profile), and how the colloid-polymer interaction strength affects colloid binding and transport. We assume that the composition of the polymer chain and of the colloid surface is homogeneous. This is the simplest possible scenario, yet it provides a rich phenomenology and enables us to formulate the basic principles that control the selectivity of colloid binding and transport *via* modulation of the solvent strength, the colloid size and the colloid-polymer interaction strength, on which more complex future models can be built.

## 2. THEORETICAL MODEL OF DIFFUSIVE COLLOID TRANSPORT ACROSS A POLYMER BRUSH

### 2.1. Defining the interaction scenario

The interaction scenario considered in this work is schematically shown in **Figure 1A**. We consider a planar polymer brush immersed in a bulk solvent with colloids, with the following parametrizations.

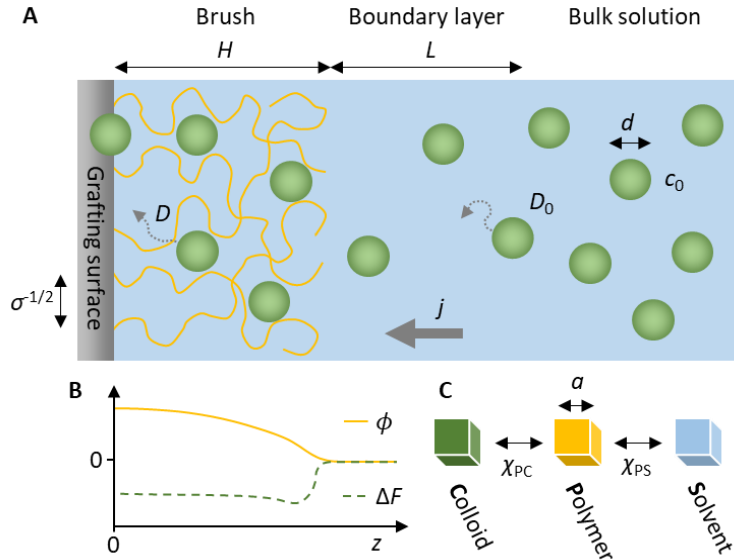

**Figure 1.** (**A-B**) Schematic illustration of the interaction scenario, and diffusive transport across a polymer brush. The brush (orange) has a thickness $H$ and a local polymer volume fraction $\phi(z)$, which depend on the grafting density $\sigma$, the degree of polymerization $N$, and the polymer-solvent Flory-Huggins interaction parameter $\chi_{PS}$. Colloid (green) interaction with the brush is quantified by the local insertion free energy $\Delta F(z)$, which depends on the colloid size $d$, the polymer-colloid interaction parameter $\chi_{PC}$, as well as $\chi_{PS}$ and $\phi(z)$. The grafting surface acts as a sink, and diffusive transport entails a flux, with flux density $j$, from the bulk solution (with colloid concentration $c_0$) to the surface. Colloid transport is influenced by the local mobility $D(z)$, which depends on $d$ and $\phi(z)$. It may be also affected by a boundary solvent layer with reduced colloid concentration compared to the bulk, with thickness $L$ depending on the experimental mass-transport conditions. (**C**) Illustration of the interaction parameters $\chi_{PS}$ and $\chi_{PC}$, with the unit length $a$ equivalent to the polymer segment size.

The polymers are linear and flexible, with segment length $a$, segment volume $a^3$, and degree of polymerization $N$, grafted with one end to a planar surface (*i.e.*, the 'grafting surface'). The surface is inert and impermeable for the polymer, and the grafting density $\sigma$ (*i.e.*, number of polymer chains per unit surface area $a^2$) is sufficiently high such that neighboring polymer chains interpenetrate and form a brush. The grafting surface is located at $z = 0$, and the brush has a thickness $H$ (all lengths, such as $z$ and $H$, are normalized by the polymer segment length $a$). The polymer brush is uniform along the surface plane, but typically exhibits a non-uniform density profile $\phi(z)$ perpendicular to the surface ($\phi$ is the polymer volume fraction; **Figure 1B**).

The colloid is approximated as a sphere with diameter $d$, as the simplest possible shape. The nearest distance of the sphere surface from the grafting surface defines the position of the colloid in or above the brush. The local colloid concentration (number density) in the brush is $c(z)$, and the colloid concentration in the bulk solvent is $c_0$.

The effects of solvent strength and colloid-polymer interactions are defined by the Flory-Huggins interaction parameters $\chi_{PS}$ and $\chi_{PC}$, with the subscripts P, S and C referring to the polymer, the solvent and the colloid, respectively (**Figure 1C**). $\chi_{PC}$ represents the difference of replacing a colloid-solvent contact by a colloid-polymer contact, and without loss of generality, we assume that the colloid-solvent interaction parameter is $\chi_{CS} = 0$. Further, we assume that the colloid concentration is sufficiently small, inside and outside the brush, such that colloid-colloid interactions can be neglected. Implicit to our approach based on Flory-Huggins interaction parameters is that polymer-polymer and polymer-colloid interactions are short-ranged, as is typically the case for neutral polymers and colloids. However, we expect our theory to remain valid also for polyelectrolyte brushes and charged colloids, provided that the ionic strength of the solution is sufficiently high such that all charge-mediated interactions are effectively short-ranged (*e.g.*, at near-physiological conditions of 150 mM monovalent ions, where the Debye length (0.8 nm) is comparable to the typical polymer segment size), and that the charges on the polymers and colloids are 'strong' such that they remain 'quenched' irrespective of the environmental conditions and binding state.

We assume the grafting surface to act as a sink for the colloids. Diffusive transport thus entails a flux of colloids from the bulk solution across the brush to the surface. The brush controls the flux density $j$ through the interaction free energy profile $\Delta F(z)$ (**Figure 1B**) and the local mobility profile $D(z)$, as described in the following sections.

## 2.2. Structure of the polymer brush

Along the lines of Ref. [20], we use the analytical strong-stretching self-consistent field (SS-SCF) approximation[18] for describing structural and thermodynamic properties of the polymer brush. As long as conformational entropy of the stretched brush-forming chains depends linearly on the extension, the self-consistent molecular potential acting on each monomer in the brush exhibits the parabolic shape,

$$\frac{\partial f[\phi(z)]}{\partial \phi(z)} = \frac{3\pi^2}{8N^2}(\Lambda^2 - z^2). \tag{1}$$

where $\Lambda$ is a constant and the free energy density of the monomer interactions in the brush $f[\phi(z)]$ can be expressed as a function of the local polymer segment concentration (volume fraction) $\phi(z)$ using the mean field Flory approximation

$$f[\phi(z)] = [1 - \phi(z)]\ln[1 - \phi(z)] + \chi_{PS}\phi(z)[1 - \phi(z)] + \phi(z)(1 - \chi_{PS}). \tag{2}$$

Here and below all the energies are normalized by the thermal energy $k_B T$. Equations (1) and (2) provide an implicit dependence of the polymer volume fraction on the distance $z$ from the grafting surface,

$$-\ln[1 - \phi(z)] - 2\chi_{PS}\phi(z) = \frac{3\pi^2}{8N^2}(\Lambda^2 - z^2), \tag{3}$$

and an expression for the interaction-related contribution to the osmotic pressure,

$$\Pi(z) = \phi(z)\frac{\partial f[\phi(z)]}{\partial \phi(z)} - f[\phi(z)] = -\ln[1 - \phi(z)] - \chi_{PS}\phi^2(z) - \phi(z). \tag{4}$$

The condition of the osmotic pressure vanishing at the edge of the brush, $\Pi(z = H) = 0$, where the local chain stretching also vanishes, provides an implicit dependence of the polymer density at the brush edge $\phi(z = H) = \phi_H$ on the Flory-Huggins parameter $\chi_{PS}$ as

$$-\ln(1 - \phi_H) - \chi_{PS}\phi_H^2 - \phi_H = 0, \tag{5}$$

which leads to an expression of the constant $\Lambda$ as

$$\Lambda^2 = H^2 - \frac{8N^2}{3\pi^2}[\ln(1 - \phi_H) + 2\chi_{PS}\phi_H]. \tag{6}$$

Notably, $\phi_H = 0$ and $\Lambda = H$ at $\chi_{PS} \leq 0.5$ (*i.e.*, in good or $\theta$-solvent), whereas $\phi_H > 0$ and $\Lambda < H$ at $\chi_{PS} > 0.5$ (*i.e.*, in poor solvent). Finally, the brush height $H$ is found as a function of $N$, $\sigma$ and $\chi_{PS}$ by using the conservation condition

$$\int_0^H \phi(z)\mathrm{d}z = N\sigma. \tag{7}$$

The evolution of the brush height and polymer density profiles upon variation of the solvent strength (controlled by the Flory-Huggins parameter $\chi_{PS}$) are illustrated in **Figure 4.**

### 2.3. Colloid insertion free energy and equilibrium partitioning in the polymer brush

We define the insertion free energy penalty $\Delta F$ as a change in free energy when a colloid particle is moved from the bulk solvent into the brush. A positive $\Delta F$ thus implies that the brush repels the colloid, and *vice versa*. As long as the colloidal particle is small in the sense that variations in polymer segment density and osmotic pressure in the brush are negligible on the $z$ scale comparable to the colloid size, the free energy change upon insertion of the colloid inside the brush can be approximated as[20]

$$\Delta F(z) = \Pi(z) \cdot V + \gamma[\phi(z)] \cdot A. \tag{8A}$$

This approximation deliberately separates the contribution of the work performed against excess osmotic pressure in the brush, which is proportional to the colloid volume $V$ (*i.e.*, the first term), from the contribution related to short-range interactions of polymer chains with the surface of the colloid (*i.e.*, the second term). The latter is proportional to the colloid surface area $A$, with the position-dependent surface tension coefficient $\gamma[\phi(z)]$. Such a separation is justified as long as the colloid effectively acts as a probe that, upon insertion into the brush, does not perturb the polymer density profile much. Equation (8A) can be generalized in a straightforward way to larger colloids by taking into account variations of the local polymer segment density and osmotic pressure across the different points on the colloid surface. For spherical colloids,

$$\Delta F(z) = \pi \int_0^d \{\Pi(z+z')z'(d-z') + \gamma[\phi(z+z')]d\}dz', \qquad (8B)$$

where $z$ is the distance from the grafting surface to the proximal point of the colloid.

Equation (4) provides an explicit expression for the local osmotic pressure $\Pi(z)$. The surface tension coefficient can be approximated as[20]

$$\gamma[\phi(z)] = (\chi_{\text{ads}} - \chi_{\text{crit}})[a_1 \phi(z) + a_2 \phi^2(z)], \qquad (9A)$$

where $\chi_{\text{ads}} \equiv \chi_{\text{PC}} - \chi_{\text{PS}}(1 - \phi(z))$ quantifies (on the mean field level) the change in the free energy upon replacement of a colloid-solvent contact by a colloid-polymer contact, which depends on the polymer-colloid interaction parameter $\chi_{\text{PC}}$, the polymer-solvent interaction parameter $\chi_{\text{PS}}$ and the polymer volume fraction $\phi(z)$. The critical value $\chi_{\text{crit}}$ assures cancellation of the conformational entropy losses arising due to spatial constraints (imposed by the impermeable colloid surface upon adsorption of a polymer segment) by the gain in the polymer-colloid contact free energy[21]. In the lattice model that implements the Scheutjens-Fleer SCF calculations the critical values is $\chi_{\text{crit}} = 6\ln(5/6) \approx -1.1$. The pre-factors $a_1$ and $a_2$ do not depend on $\chi_{\text{PC}}$ and $\chi_{\text{PS}}$, to a first approximation. Here, we use the values $a_1 = 0.18$ and $a_2 = -0.09$, as obtained in our previous work[20] based on the fitting of the results of numerical Scheutjens-Fleer calculations by the analytical Equation (3).

The surface tension coefficient $\gamma$ can be given a simple interpretation when considering the regime of relatively low polymer densities. Neglecting the quadratic term in Equation (9A), the excess free energy of one monomer-surface contact can be estimated as

$$\varepsilon \approx a_1(\chi_{\text{ads}} - \chi_{\text{crit}}). \qquad (9B)$$

In our numerical calculations, the polymer-colloid interaction parameter is varied in the range $0 \geq \chi_{\text{PC}} \geq -3$, while the polymer-solvent interaction parameter is varied in the range $0 \leq \chi_{\text{PS}} \leq 1$. Hence the range of values for the free energy per contact explored here can be estimated as $0.20 > \varepsilon > -0.52$, *i.e.*, below thermal energy $k_\text{B}T$.

At equilibrium, the local concentration of colloids in the brush is proportional to the Boltzmann factor provided their concentration is small enough to neglect their mutual interactions. The pre-factor is determined by the concentration in the bulk solvent, $c_0$, such that

$$c_{\text{eq}}(z) = c_0 e^{-\Delta F(z)}. \tag{10}$$

The ratio between the average colloid concentration in the brush at equilibrium, $\langle c_{\text{eq}} \rangle$, and the bulk concentration, defines the average partition coefficient

$$\frac{\langle c_{\text{eq}} \rangle}{c_0} = \frac{1}{H} \int_0^H e^{-\Delta F(z)} dz, \tag{11}$$

which can be measured experimentally and may be used to test the theoretical description of the brush/solvent/colloid system in terms of equilibrium colloid binding properties.

### 2.4. Effects of the polymer mesh on colloid diffusion

Locally, the brush structure can be described as a semi-dilute polymer solution with the segment concentration $\phi(z)$. A semi-dilute solution is characterized by a concentration-dependent correlation length $\xi$ which has the meaning of a typical mesh size formed by overlapping polymer coils.

A scaling theory of colloid diffusion in semi-dilute solutions was proposed by Cai *et al.* [22]. Two scaling regimes were identified that are relevant in our context, when the colloid size is below the polymer entanglement length. Colloids smaller than the correlation length are not hindered by the polymer and their diffusion coefficient is the same as in the pure bulk solvent and is given by the Stokes-Einstein formula

$$D_0 \simeq \frac{k_B T}{\eta_S d} \quad \text{for} \quad d < \xi, \tag{12A}$$

where $\eta_S$ is the solvent viscosity. The approximately equal sign here indicates that any numerical pre-factors of order unity have been dropped in the scaling approximation.

On the other hand, colloids larger than the mesh size undergo hopping diffusion: they are temporarily trapped within the mesh cage and must wait for the polymer chains to relax to hop from one local polymer network cage to another. As a result, the diffusion is slowed down so that the long-term diffusion in the polymer mesh is given by [22]

$$D_m \simeq D_0 \frac{\xi^2}{d^2} \quad \text{for} \quad d > \xi. \tag{12B}$$

We use a simple expression that interpolates between Equations (12A-B) by assuming that the retarding effects due to friction against solvent and mesh relaxation are additive, and take the numerical pre-factor in Equation (12B) to be equal to 1,

$$D^{-1}(z) = D_0^{-1}[1 + d^2/\xi^2]. \tag{13A}$$

In the following discussion of colloid transport across a brush, we account for this effect by introducing a position-dependent diffusion coefficient $D(z)$ characterizing the local mobility of the colloid. This depends on the local correlation length $\xi(z)$ which, in turn, is determined by the local polymer segment concentration $\phi(z)$. The scaling relation between the correlation length and the polymer concentration depends on the solvent quality[23]. We are exploring a broad range of values for the Flory-Huggins parameter $\chi_{PS}$, from good ($\chi_{PS} = 0$) to rather poor ($\chi_{PS} = 1$) solvents. In order not to further complicate the theory we use $\xi = \phi^{-1}$, which is characteristic of the $\theta$–solvent ($\chi_{PS} = 0.5$), eventually leading to a position-dependent local mobility of the form

$$D(z) = \frac{D_0}{1+\phi^2(z)d^2}. \tag{13B}$$

**Figure 2** illustrates the smooth interpolation between Equations (12A-B) by Equation (13B). We note that our numerical results are not very sensitive to the particular choice of the scaling expression for $\xi(\phi)$.

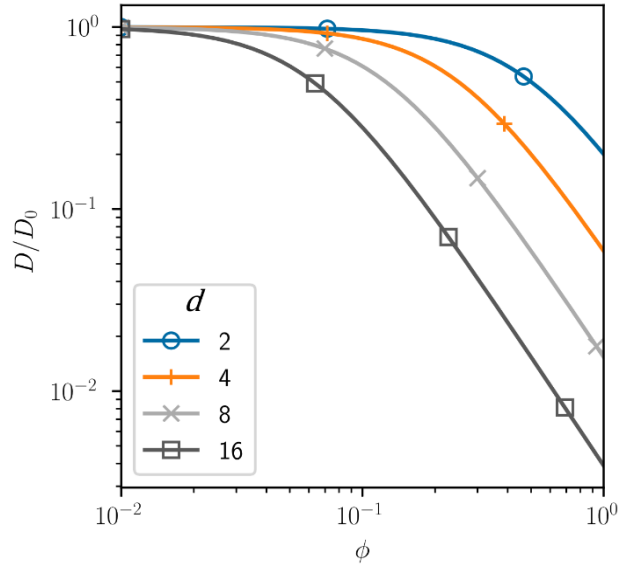

**Figure 2.** Effect of polymer concentration on colloid mobility inside the polymer brush. The colloid mobility $D$, normalized by the Stokesian diffusion coefficient $D_0$, is shown as a function of the polymer volume fraction $\phi \sim \xi^{-1}$, for a range of colloid sizes ($d = 2, 4, 8$ and $16$; symbols and color codes as indicated).

It should be noted that the theory by Cai *et al.*[22] assumes that the diffusing colloids are non-sticky. We are exploring both repulsive and attractive regimes of the polymer-colloid interactions. However, even in the case of relatively strong adsorption studied in this work an individual contact (*i.e.*, of a polymer strand with the size comparable to the segment length $a$) has a free energy smaller in magnitude than $k_B T$ (since $|\varepsilon| < 1$), as estimated above based on Equation (9B). As we assume the colloid surface to be homogeneous, the activation energy controlling the individual contact dynamics does not grow with the size of the colloid, in contrast to the insertion free energy which is eventually responsible for the most dramatic effects on the colloid transport (*vide infra*). Although colloid-polymer attraction can bring some additional slowing-down, the subject is studied little and we do not account for this effect explicitly.

### 2.5. Stationary colloid flux across the polymer brush

For colloid transport from the bulk solution into and through a brush, two important factors must be taken into account. Firstly, the gradient of insertion free energy generates a force that affects the motion of the diffusing colloid. Secondly, even if the free energy profile were perfectly flat, the mobility of a colloid within the interior of the brush would be affected by the presence of the brush-forming chains as discussed above.

Diffusion of colloidal particles in the presence of an effective potential is described by the Smoluchowsky equation which represents a high-friction limit of the Fokker-Planck equation[24]. In our case, the brush is laterally homogeneous so that the only relevant coordinate is $z$, and the equation simplifies to

$$\frac{\partial c(z,t)}{\partial t} = \frac{\partial}{\partial z} D(z) \left( \frac{\partial c(z,t)}{\partial z} + c(z,t) \frac{\partial \Delta F(z)}{\partial z} \right), \tag{14}$$

where $c(z,t)$ is the local colloid concentration, $D(z)$ the local colloid mobility (Equation (13)), and $\Delta F(z)$ the position-dependent colloid insertion free energy (Equation (8)).

Analysis of the colloid transport through a brush can take various directions. In particular, the translocation problem focuses on the mean first-passage time which is appropriate when colloids can be transported only consecutively, one by one. In the case of a planar brush this mechanism does not reflect potential experimental conditions unless one considers the limit of vanishing concentration of colloids so that at any moment no more than one colloid is present. When a multitude of elementary transport processes take place in parallel, the analysis of the flux density under stationary conditions seems more appropriate, although in the case of a single brush (asymmetric insertion free energy profile) both approaches give very similar results. Here, we implement the stationary flux approach since it yields a simpler interpretation. The flux density has the form

$$j(z,t) = -D(z) \left[ \frac{\partial c(z,t)}{\partial z} + c(z,t) \frac{\partial \Delta F(z)}{\partial z} \right]. \tag{15}$$

The stationarity condition means that both the colloid flux and the colloid density are time-independent, which by the continuity equation implies that the flux density is the same at every point in space, $j(z) = j_0 = \text{const}$. The solution of the stationary version of Equation

(15) with non-zero flux is sought in the form of a modified Boltzmann distribution $c(z) = \psi(z)e^{-\Delta F(z)}$. It follows that the function $\psi(z)$ must satisfy the equation

$$\frac{\partial \psi(z)}{\partial z} = j_0 D^{-1}(z) e^{\Delta F(z)}. \tag{16}$$

Here and below, $j_0$ has the meaning of the absolute value of the diffusive flux density, while its direction is along the negative $z$ axis since the source of the colloids is located at $z > 0$, *i.e.*, above the grafting plane which serves as an imperfect sink. The general solution of Equation (16) is given by

$$\psi(z) = B + j_0 \int_0^z D^{-1}(z') e^{\Delta F(z')} dz', \tag{17}$$

where $B$ is the integration constant. We consider the situation where the brush is grafted onto a substrate that can absorb the colloids with a certain probability. This is accounted for by a mixed boundary condition $c(z = 0) = r j_0$, where $r$ is a positive constant. $r = 0$ describes a perfect sink, while $r \to \infty$ corresponds to an impenetrable reflecting surface. The other boundary condition is imposed at a distance $z = H + L$ from the grafting surface where the source of the colloids is located, $c(H + L) = c_0$. This distance includes the thickness of the brush, $H$, and a boundary layer of thickness $L$ within the solvent defined by the actual conditions of the experiment to maintain a constant bulk colloid concentration, see **Figure 1A**. In technologically relevant conditions this could be accomplished, for example, through shear flow in a microfluidic channel with brush-coated walls[25]; in biological systems (*e.g.*, inside cells), colloids (*e.g.*, proteins) may be replenished locally through their enzymatic production. In the following, we show that the transport properties of the brush itself can be characterized independently of the details of the boundary layer.

Using the general solution (Equation (17)), the boundary conditions, and the fact that $\Delta F(z) = 0$ and $D(z) = D_0$ beyond the outer brush edge ($z > H$), one obtains a relation between the concentration at the source, $c_0$, and the stationary flux density, $j_0$, as

$$c_0 = j_0 \left[ \int_0^H D^{-1}(z) e^{\Delta F(z)} dz + D_0^{-1} L + r e^{\Delta F(0)} \right]. \tag{18}$$

Finally, the stationary concentration profile is expressed as

$$c(z) = c_0 e^{-\Delta F(z)} \frac{r e^{\Delta F(0)} + \int_0^z D^{-1}(z') e^{\Delta F(z')} dz'}{r e^{\Delta F(0)} + \int_0^H D^{-1}(z') e^{\Delta F(z')} dz' + D_0^{-1} L}. \tag{19}$$

### 2.6. Brush permeability

The result of Equation (18) admits a simple interpretation since it has the form of Ohm's law for resistors connected in series. Indeed, $c_0$ plays the role of the voltage applied, $j_0$ is the current density, and the three terms in square brackets can be identified as the resistances of the brush, $R_{\text{brush}} = \int_0^H D^{-1}(z)e^{\Delta F(z)}dz$, of the boundary solvent layer, $R_{\text{sol}} = D_0^{-1}L$, and of the imperfect sink, $R_{\text{sink}} = re^{\Delta F(0)}$, respectively. Clearly, $D_0^{-1}$ plays the role of the resistivity attributed to the pure solvent (which is naturally multiplied by the thickness of the depletion layer $L$), while the integrand in the brush term, $D^{-1}(z)e^{\Delta F(z)}$, represents the local resistivity within the brush that accounts both for the slowed-down diffusion affected by the polymer mesh and for the insertion free energy; integration over the brush thickness represents a sum of the resistances of infinitesimally thin layers.

Since the three different contributions to the total resistance are additive, the transport properties of the brush can be analyzed separately, irrespective of the experimental geometry, and of the properties of the imperfect sink. The permeability of the brush for diffusive colloid transport is defined through

$$P = \frac{H}{R_{\text{brush}}} = \left[\frac{1}{H}\int_0^H D^{-1}(z)e^{\Delta F(z)}dz\right]^{-1}. \tag{20}$$

The permeability of a brush, as defined in Equation (20), effectively is a material parameter, analogous to the permeability of conventional separation membranes. Of note, $P$ would reduce to $D_0$ if the brush were completely 'invisible' (transparent) to the colloidal particle.

A permeability parameter more directly related to the experimental setup would include all the contributions defining the net diffusive flux

$$P_{\text{net}} = \left\{\frac{1}{H+L}\left(\int_0^H D^{-1}(z)e^{\Delta F(z)}dz + D_0^{-1}L + re^{\Delta F(0)}\right)\right\}^{-1}. \tag{21}$$

In the following we mostly focus on the properties of the brush; to make the analysis more transparent we consider the case of an ideal sink ($r = 0$).

## 2.7. Impact of the brush-solution interface on colloid partitioning and permeability

The two quantities in the focus of our interest are the average partition coefficient, $\langle c_{\text{eq}}\rangle/c_0 = \frac{1}{H}\int_0^H e^{-\Delta F(z)}dz$ (Equation (11)), and the brush permeability (Equation (20)) or, alternatively, the brush resistance, $R_{\text{brush}} = \int_0^H D^{-1}(z)e^{\Delta F(z)}dz$. The relevant expressions have a similar structure that involves integration of an exponential term containing the insertion free energy $\Delta F$. Here, we aim to clarify the important role of the interface between the brush and the bulk solution in determining the two quantities of interest.

Since we define the position of the colloid by its nearest distance to the grafting surface, the colloid with coordinates in the range $0 < z \leq H - d$ is completely immersed into the brush while the range $H - d < z < H$ corresponds to a partially immersed colloid. Within the region of partial immersion (*i.e.*, the interface region) the insertion free energy profile demonstrates a crossover from $\Delta F = 0$ in the bulk solution ($z \leq H$) to the finite values – positive for repulsive brushes, and negative for attractive brushes – characteristic of the brush interior. For relatively large colloids the interface region not only constitutes a non-negligible fraction of the total brush thickness. More importantly, $\Delta F$ becomes large in absolute values, and one or the other of the $e^{\pm \Delta F(z)}$ terms in Equations (11) and (20) may become very small, depending on the sign of $\Delta F$.

Cartoons with simplified insertion free energy profiles, and the corresponding Boltzmann weight ($e^{-\Delta F(z)}$) and inverse Boltzmann weight ($e^{\Delta F(z)}$) profiles, are presented in **Figure 3** for a repulsive ($\Delta F > 0$) and an attractive ($\Delta F < 0$) brush. For the sake of order-of-magnitude estimates we ignore the full complexity of the $\Delta F$ profile, and instead assume that $\Delta F$ simply interpolates linearly in the interface region (*i.e.*, as the colloid becomes increasingly immersed) from $\Delta F = 0$ in the bulk solution to a constant (typical) value ($\Delta F = \Delta F^*$) in the brush interior.

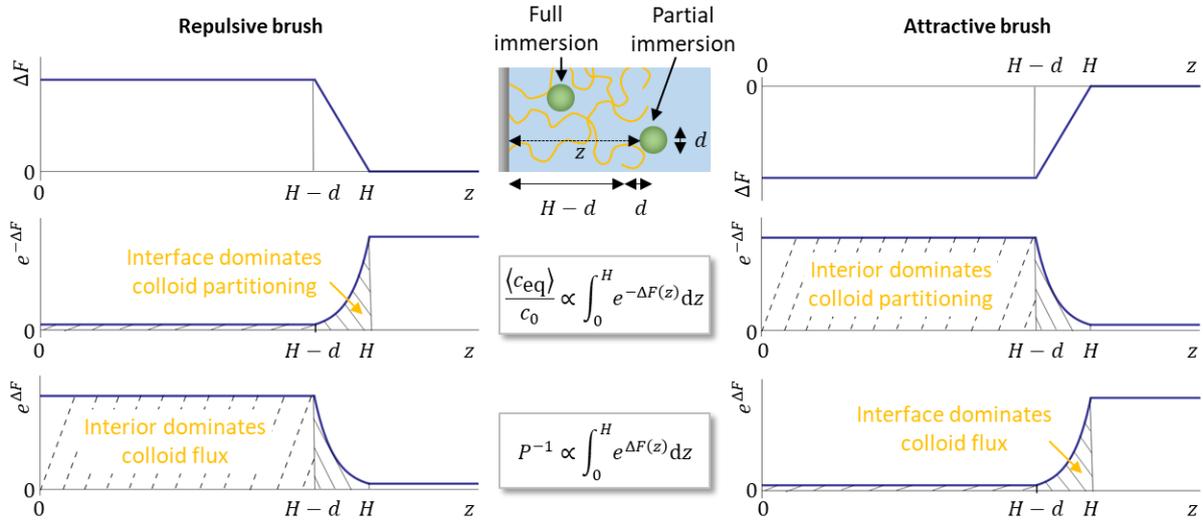

**Figure 3.** Cartoons illustrating the insertion free energy profile $\Delta F(z)$ (**top**), the Boltzmann weight profile $e^{-\Delta F(z)}$ (**middle**), and the inverse Boltzmann weight profile $e^{\Delta F(z)}$ (**bottom**) for a repulsive brush ($\Delta F > 0$; **left**) and an attractive brush ($\Delta F < 0$; **right**). The average partition coefficient is proportional to the area under the Boltzmann weight profile, while the brush resistance is proportional to the area under the inverse Boltzmann weight profile. Contributions due to the brush interior and the interface, respectively, are shown by distinct dashes. Depending on the sign of $\Delta F$, the colloid partitioning and the flux are dominated either by the brush interior or by the interface (as indicated).

Considering first the case of a repulsive brush ($\Delta F > 0$), it is clear that the Boltzmann weight defining the average partition coefficient, $e^{-\Delta F(z)}$, is very small in the brush interior. Under our simplifying assumptions, the contribution of the brush interior to the partition coefficient is $\frac{1}{H}\int_0^{H-d} e^{-\Delta F^*} dz = \frac{H-d}{H} e^{-\Delta F^*}$, while the contribution from the interface region is $\frac{1}{H}\int_{H-d}^{H} e^{-\Delta F(z)} dz = \frac{1}{H}\int_{0}^{d} e^{-\Delta F^*(1-z/d)} dz \approx \frac{d}{H\Delta F^*}$. The ratio of the two contributions is $\frac{H-d}{d}\frac{\Delta F^*}{e^{\Delta F^*}}$; it follows that, for sufficiently large $\Delta F^*$, the average partition coefficient is dominated by the interface contribution (since $e^{\Delta F^*}$ increases much more rapidly than $\Delta F^*$; $\langle c_{eq}\rangle/c_0 \approx \frac{d}{H\Delta F^*}$). The relative importance of the two contributions is illustrated in **Figure 3** (left) by the dashed areas under the corresponding portions of the $e^{-\Delta F(z)}$ curve.

On the other hand, the inverse Boltzmann weight defining the brush resistance to colloid flow, $e^{\Delta F(z)}$, is generally large, and the contributions due to the brush interior and interface can be estimated as $D^{-1} e^{\Delta F^*}(H-d)$ and $D^{-1} e^{\Delta F^*}\frac{d}{\Delta F^*}$, respectively. Neglecting any variations in the colloid mobility $D$ between the brush interior and interface, the ratio of the two contributions here scales as $\Delta F^*(H-d)/d$, implying that the brush interior dominates the brush resistance and permeability, with the brush interface only making a minor contribution ($R_{brush} \approx \frac{e^{\Delta F^*}H}{D}$ and $P \approx \frac{D}{e^{\Delta F^*}}$).

In the case of an attractive brush, the insertion free energy is negative ($\Delta F < 0$), and the two exponential profiles switch their shapes (**Figure 3**). Correspondingly, the importance of the interface contributions to the partition coefficient and to the brush resistance is also reversed. Now, it is the brush resistance that is dominated by the interface, while the interior contribution is exponentially small ($R_{brush} \approx \frac{d}{D|\Delta F^*|}$ and $P \approx D|\Delta F^*|\frac{H}{d}$). Conversely, the average partition coefficient for an attractive brush is dominated by the brush interior, and the interface provides only a minor correction ($\langle c_{eq}\rangle/c_0 \approx e^{|\Delta F^*|}$).

If the brush density profile were uniform, and interface effects completely absent, then the integrands in Equations (11) and (20) would be constants and integrations trivial. Under these simplistic conditions, we obtain a simple relation between the permeability and the partition coefficient, $P = D\langle c_{eq}\rangle/c_0$, which is well known and commonly used in membrane science. Clearly, neglecting the interfacial effects is justified if the magnitude of the insertion free energy is small ($|\Delta F| \lesssim 1$) and/or the thickness of the membrane is much larger than the size of the diffusing particle ($H \gg d$). These conditions are readily satisfied, for example, in conventional gas separation membranes. In contrast, polymer brushes can be seen as a special type of membrane: that brushes are usually not much thicker than colloids are large, and exhibit a non-uniform polymer density profile, makes the relation between the permeability and the average partition coefficient more subtle.

## 3. RESULTS

### 3.1. Conformation of a representative polymer brush

To illustrate the effects of the brush and colloid properties on colloid permeability and partitioning, we selected a brush with polymers made from $N = 2000$ segments and grafted at a density of $\sigma = 0.08$. The latter is equivalent to a root-mean-square distance of 3.5 segment lengths between neighboring anchor points. The thickness $H$ and the average polymer volume fraction $\langle \phi \rangle = N\sigma/H$ as a function of the solvent quality $\chi_{PS}$ were derived *via* Equation (7) and are shown in **Figure 4A**. The brush thickness varies over a wide range across the $\chi_{PS}$ range covered, from 765 in good solvent ($\chi_{PS} = 0$) *via* 462 in $\theta$-solvent ($\chi_{PS} = 0.5$) to 228 in a relatively poor solvent ($\chi_{PS} = 1$), the latter being close to the lower thickness limit for a solvent free brush ($H = N\sigma = 160$). Equivalently, the average polymer volume fraction ranges from 0.21 in good solvent ($\chi_{PS} = 0$) to 0.7 in poor solvent ($\chi_{PS} = 1$).

The thickness of the most compact brush (160) exceeds the root-mean-square distance between anchor points ($\sigma^{-1/2} \approx 3.5$) by far, guaranteeing in-plane homogeneity of the brush at any solvent strength. Importantly, the brush thickness also remains substantially larger than the largest colloids considered ($d = 16$), implying that all colloids are readily immersed within the brush.

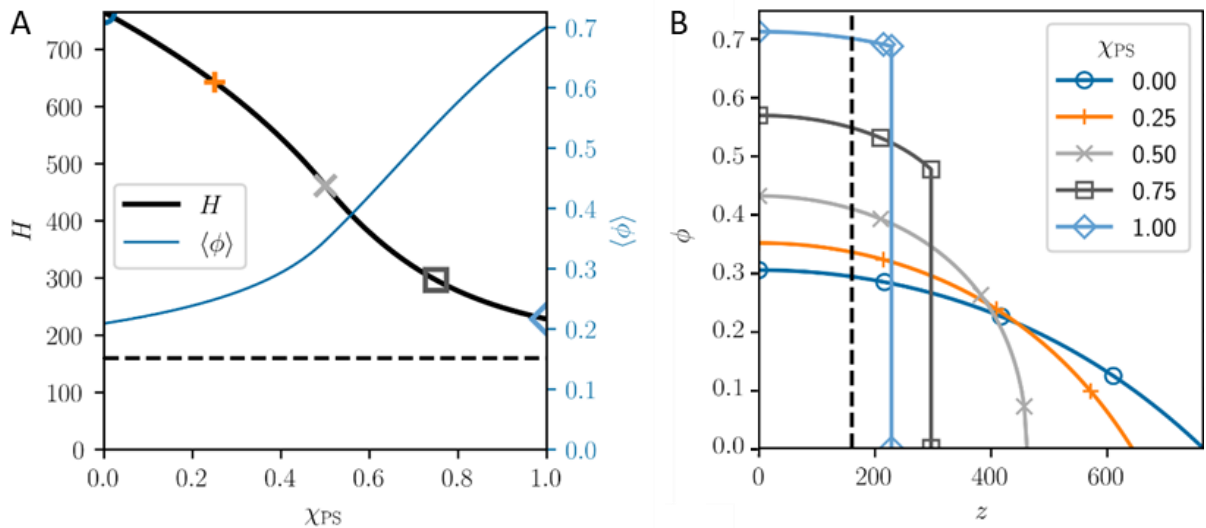

**Figure 4.** Structural properties of a representative polymer brush ($N = 2000$ and $\sigma = 0.08$) as a function of the solvent quality $\chi_{PS}$. (**A**) Brush thickness $H$ (thick solid black line, left axis) and average polymer volume fraction $\langle \phi \rangle$ (thin solid blue line, right axis). (**B**) Polymer volume fraction profile for $\chi_{PS} = 0, 0.25, 0.5, 0.75$ and 1 (as indicated in the legend; the same symbols are shown in panel **A** at the corresponding $\chi_{PS}$ levels). The dashed horizontal and vertical lines indicate the thickness $N\sigma$ of a compact, solvent free brush.

**Figure 4B** illustrates the expected polymer density profiles: profiles are parabolic in good solvent ($\chi_{PS} < 0.5$) and are truncated with a sharp brush boundary in poor solvent ($\chi_{PS} > 0.5$), in good agreement with earlier work[18, 19].

### 3.2. Brush permeability to colloids - Mapping the parameter space

**Figure 5** presents contour plots of the permeability $P$ (normalized by the Stokesian diffusion coefficient in pure solvent $D_0$) as a function of the solvent strength $\chi_{PS}$ (from good solvent, $\chi_{PC} = 0$, to poor solvent $\chi_{PC} = 1$) and the polymer-colloid affinity parameter $\chi_{PC}$ (from inert, $\chi_{PC} = 0$, to strongly attractive, $\chi_{PC} = -3$) for a set of colloid sizes (from small, $d = 2$, to large, $d = 16$, compared to the mean polymer anchor spacing, $\sigma^{-1/2} \approx 3.5$).

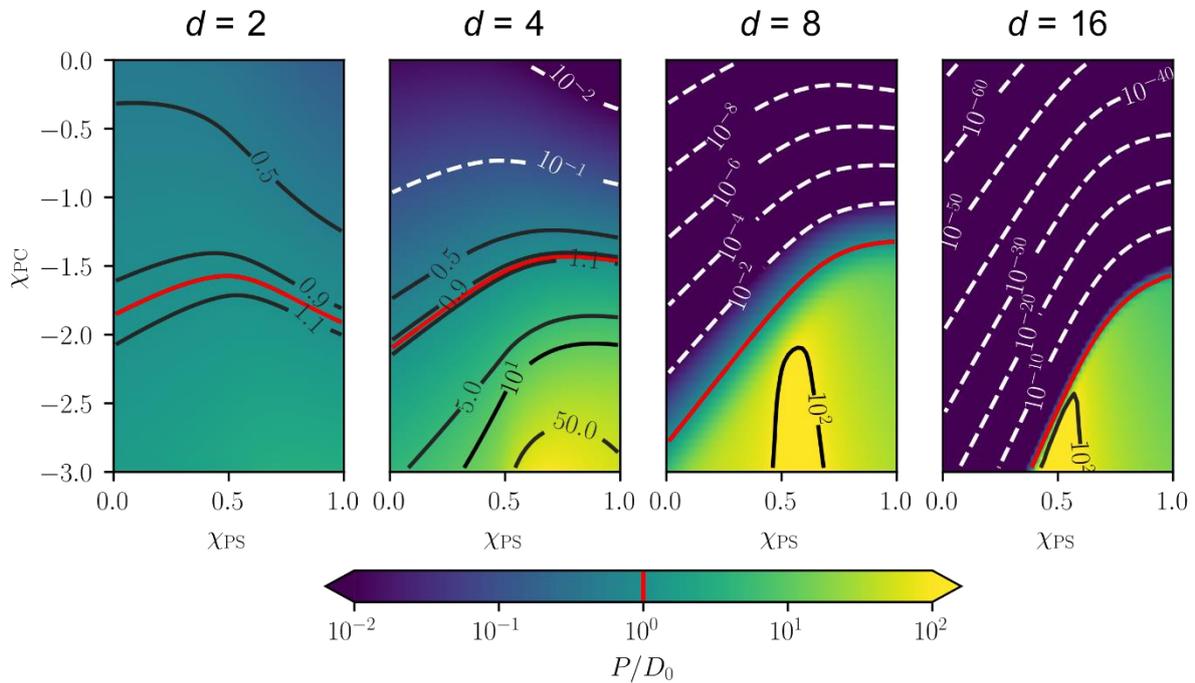

**Figure 5.** Effect of solvent strength and polymer-colloid affinity on the brush permeability to colloids. Contour plots for the normalized permeability $P/D_0$ for a range of colloid sizes ($d = 2, 4, 8$ and $16$, as indicated atop the graphs) as a function of the interaction parameters $\chi_{PS}$ and $\chi_{PC}$. The solid red lines indicate $P/D_0 = 1$. Brush parameters: $N = 2000$, $\sigma = 0.08$.

The colloid transport through a polymer brush can be strongly impeded or facilitated in comparison to the solution layer of the same thickness. The most important factor (especially for colloids of larger size) is the maximal value of the colloid insertion free energy profile, $\max(\Delta F)$, which dominates the exponential factor in the integral defining the brush resistance. One can identify two qualitatively distinct regimes: a repulsive brush with $\max(\Delta F) > 0$ (impeded permeation), and an attractive brush with $\max(\Delta F) < 0$ (facilitated permeation). A secondary factor that always leads to transport slowing down is the reduced colloid mobility within the brush due to the polymer mesh effect. The insertion free energy is controlled by three parameters: the solvent quality for the brush-forming polymer, the polymer-colloid affinity, and the colloid size. In order to have strongly repulsive and strongly attractive brush regimes, the colloid size must be larger than the polymer Kuhn segment length by about an order of magnitude or more.

Good solvent conditions typically produce a repulsive brush, mostly due to high osmotic pressure. The attractive brush regime requires a combination of preferably poor solvent conditions and of pronounced adsorption of the polymer onto the surface of the colloid (*i.e.*, the magnitude of the affinity parameter well exceeding the critical value, $|\chi_{PC}| > |\chi_{crit}|$). Strongly repulsive brushes have exponentially low permeability ($P \sim e^{-\max(\Delta F)}$). Strongly attractive brushes exhibit enhanced permeability which is mostly determined by the brush-solution interface as explained in Section 2.7.

One would expect the maximum of permeation in the right lower corner of each contour plot corresponding to the lowest solvent strength and highest polymer-colloid affinity, where $\max(\Delta F)$ is the lowest. **Figure 5**, however, shows a non-monotonic dependence of permeation on solvent strength with an optimum value of $\chi_{PS}$ slightly above $\theta$-solvent. The main reason behind this effect is that the average polymer density of the brush increases with decreasing solvent quality (**Figure 4**) leading to a reduced colloid mobility. This effect is particularly pronounced for large particles.

### 3.3. Permeability of polymer brushes - Effect of colloid size

**Figure 6** shows the brush permeability as a function of colloid size, for a set of solvent strengths (from good solvent, $\chi_{PS} = 0$, *via* $\theta$-solvent, $\chi_{PS} = 0.5$, to poor solvent, $\chi_{PS} = 1$) and colloid interaction strengths (from inert, $\chi_{PC} = 0$, to strongly attractive, $\chi_{PC} = -3$). We discriminate regimes of facilitated and impeded permeation by comparison to the permeability $D_0 \sim d^{-1}$ of the bulk solvent (dashed black lines in **Figure 6**). Four distinct regimes can be discerned in the dependence of permeability on colloid size; these are best appreciated for $\chi_{PS} = 0.5$ and $\chi_{PC} = -3$, and numbered I to IV in **Figure 6**.

In **Regime I**, the magnitude of the insertion free energy is small (close to $k_B T$ or less) and the colloid mobility in the brush is comparable to the bulk solution; the permeability thus remains close to $D_0$.

In **Regime II**, the insertion free energy increases in magnitude and is dominated by the attractive interaction of the colloid with the brush interior ($\Delta F \sim -d^2$, see Equation (8)), whereas the colloid mobility is only moderately affected by the brush. As a consequence, the permeability increases strongly.

In **Regime III**, the permeability is dominated by the solvent-brush interface and according to simple estimates presented in Section 2.7 ($P \approx D|\Delta F^*|\frac{H}{d}$) we expect the permeability to decrease with colloid size. This regime is realized in attractive brushes for sufficiently high colloid affinity and large colloid size. The switch between regimes II and III corresponds to the maximal possible permeation for a given $\chi_{PS}$ and $\chi_{PC}$.

The transition to **Regime IV** happens when the faster growing ($\sim d^3$) repulsive osmotic contribution to the insertion free energy overcomes the attractive surface contribution which grows slower ($\sim d^2$). As a result the insertion free energy changes sign and the brush quickly becomes repulsive for larger colloids leading to a dramatic drop in permeability.

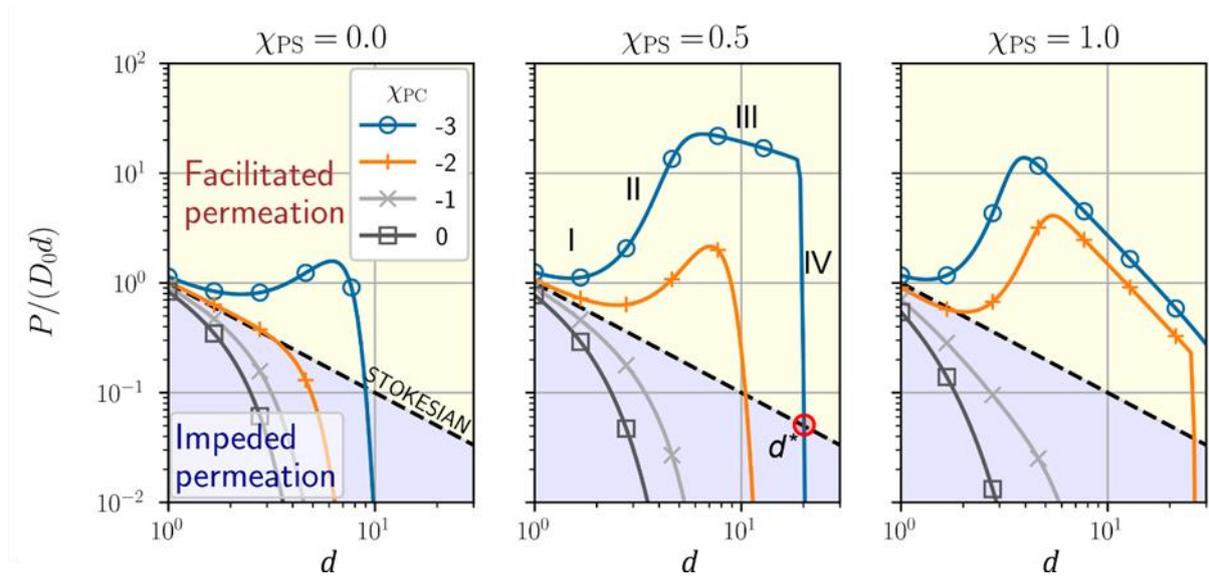

**Figure 6.** Effect of colloid size on the brush permeability to colloids. Plots of the permeability $P$ as a function of colloid size $d$ for a set of solvent strengths ($\chi_{PS} = 0, 0.5$ and $1.0$; as indicated atop the graphs) and colloid interaction strengths ($\chi_{PC} = 0, -1, -2, -3$; with symbols and colors as indicated in the legend). $P$ is normalized by $D_0 d = k_B T/\eta_S$ (see Equation (10A)); the dashed black lines indicate Stokesian diffusion coefficient ($P = D_0$) and separate the regimes of facilitated and impeded permeation relative to bulk solvent. Four distinct permeability regimes (I to IV), and the critical colloid size $d^*$ (red circle) are highlighted in the middle panel for $\chi_{PS} = 0.5$ and $\chi_{PC} = -3$.

The extent to which the four regimes are expressed depends sensitively on the polymer-colloid affinity and solvent strength. The attractive, interface-dominated regime III, for example, is effectively absent in good solvent yet more extended in poor solvent. Naturally, the attractive regimes II and III are absent for inert or slightly attractive colloids ($\chi_{PC} \geq -1$), and instead there is a more gradual transition from the (weakly repulsive) Stokesian regime I to the strongly repulsive regime IV.

In an experimentally relevant situation, the colloid flux is affected not only by the brush properties but also by the thickness of the adjacent boundary solution layer, similar to the situation when the net current is controlled by two resistors connected in series (see Section 2.5). For brushes that facilitate permeation ($P > D_0$) and thus have a low resistance, the boundary layer can limit the net colloid flux $P_{\text{net}}$ substantially. To realize enhanced transport through attractive brushes the thickness $L$ of the boundary layer should be kept small (below the micrometer range in our example; see **Figure S1**). We note that biological cells provide a natural setting for enhanced transport scenarios owing to the small intercellular distances. Repulsive brushes are much less susceptible to boundary layer effects due their high intrinsic resistance.

As far as permselectivity with respect to colloid size is concerned the crossover from Regime III to Regime IV is the most promising and robust since it implies effective gating of all the colloids that are larger than the threshold size $d^*$ corresponding to the critical condition $P = D_0$. The critical condition implies that the brush becomes transparent to colloids since its permeability matches that of a pure solvent. In the next section, we analyze the near-critical behavior and gating around $P = D_0$ in more detail.

### 3.4. Permselectivity and gating in the near-critical regime

The polymer-colloid interaction strength $\chi_{\text{PC}}$ affects the insertion free energy in the most fundamental way: by tuning its value into the adsorption regime (considerably below the critical value $\chi_{\text{crit}}$) one can change the sign of the maximum insertion free energy. The conditions when the maximum insertion free energy is close to zero turns out to be most beneficial for the brush to display simultaneously high selectivity and high permeability. We will refer to these conditions as near-critical. The sign of the maximum insertion free energy is determined by the competition between the two terms in Equation (8A): the volume contribution is always positive, while the surface contribution in the adsorption regime is negative. Here we present order-of-magnitude estimates to understand the selectivity of the brush operating in the near-critical regime. In order to simplify the analysis we use the box model of the brush which assumes a uniform density profile and thus the insertion free energy $\Delta F$ is independent of $z$. In poor solvents, the real brush density is almost uniform and the box model approximation is well justified. Moreover, we keep aside the polymer mesh effect leading to a reduced colloid mobility and focus only on the behavior of the exponential factor as the more important one for larger particles. Within this simplified approach the critical condition reduces to $\Delta F = 0$. For spherical colloids at the critical condition, Equation (8A) becomes

$$\frac{\pi}{6}\Pi d^3 + \gamma \pi d^2 = 0. \tag{26}$$

Here $\Pi(\chi_{PS})$ is the osmotic pressure controlled by the solvent quality, and $\gamma(\chi_{PC}, \chi_{PS})$ is the surface tension which is directly controlled by the polymer-colloid affinity and indirectly by the solvent quality. It is clear that the critical condition calls for $\gamma(\chi_{PC}, \chi_{PS}) < 0$ which requires strong enough polymer-colloid affinity.

The simplified critical condition (Equation (26)) can be achieved by changing either the colloid properties ($d$ and $\chi_{PC}$) or the solvent quality ($\chi_{PS}$). It is clear that for large enough particles the brush is invariably repulsive because the positive volume term grows faster with $d$ than the negative surface term. Equation (26) leads to a simple expression for the critical diameter, at which the volume and surface terms cancel each other,

$$d^* = -\frac{6\gamma}{\Pi}. \tag{27}$$

For example, at $\chi_{PS} = 1$ (poor solvent) and $\chi_{PC} = -2$, the brush density is $\phi \approx 0.7$ which gives the estimates $\gamma \approx -0.1$, $\Pi \approx 0.03$, and $d^* \approx 20$. Beyond this diameter one expects a very sharp decrease in the brush permeability (Regime IV in **Figure 6**). It is clear that $d^*$ grows with the adsorption strength which enters linearly in $\gamma$ (see Equation (8B)) and also grows with the increase in the Flory-Huggins parameter $\chi_{PS}$, since the brush osmotic pressure goes down in poor solvents while the surface tension decreases (*i.e.*, increases in magnitude).

On the other hand, for a colloid of a given diameter the critical condition may be achieved by increasing the adsorption strength or by tuning the solvent quality. In the vicinity of the critical condition, we Taylor expand the insertion free energy with respect to the three control parameters

$$\Delta F \approx \frac{\pi}{6}\Pi^* d^{*2}(d - d^*) + 0.57\phi^* d^{*2}(\chi_{PC} - \chi_{PC}^*) - 0.1|\chi_{PC}^*|^{1.5} d^{*2}(\chi_{PS} - \chi_{PS}^*). \tag{28}$$

Here the colloid diameter, the osmotic pressure, the average brush density and the interaction parameters evaluated at the critical condition (27) are indicated by asterisks. The pre-factors appearing with the $(\chi_{PS} - \chi_{PS}^*)$ term in Eq (28) are not derived rigorously but rather indicate the typical values of the partial derivatives of the osmotic pressure and the surface tension with respect to the solvent quality in the relevant range of the interaction parameters $\chi_{PC}$ and $\chi_{PS}$.

Importantly, all the partial derivatives are proportional to the square of the critical colloid diameter which makes the insertion free energy very sensitive to all three parameters in the near-critical regime for large enough colloids. Once the brush shifts into the repulsive regime, its permeability is reduced exponentially by a factor of order $e^{-\Delta F}$. Thus, if the critical colloid size is large ($d^* \geq 8$), the brush permeability can be qualitatively described as a step function, and transport is blocked for colloids that either have an insufficient affinity for the polymer, or are too large. As an example, for the parameters discussed above ($\gamma \approx -0.1$, $\Pi \approx 0.03$, $d^* \approx 20$) an increase in the particle size by one segment length, from $d^* = 20$ to $d = 21$, would result in the insertion free energy change from $\Delta F \approx 0$ to $6$ with a corresponding drop in permeability by more than two orders of magnitude. Equation (28) also implies that the near-critical condition, and therefore the gating behavior with respect to the colloid parameters (size and affinity to polymer), can be tuned by changing the solvent quality.

The gating effect with respect to the polymer-colloid affinity can be clearly seen in the two right panels of **Figure 7** displaying the results of our numerical analysis for the detailed brush model with a non-uniform density profile. Gating is very sharp for $d \geq 8$. Weaker and more continuous effects are observed for smaller particles ($d = 4$), and for the smallest particles ($d = 2$) the brush is effectively transparent, because the insertion free energy is smaller than $k_B T$ and the colloid mobility in the brush is essentially the same as in the bulk solution.

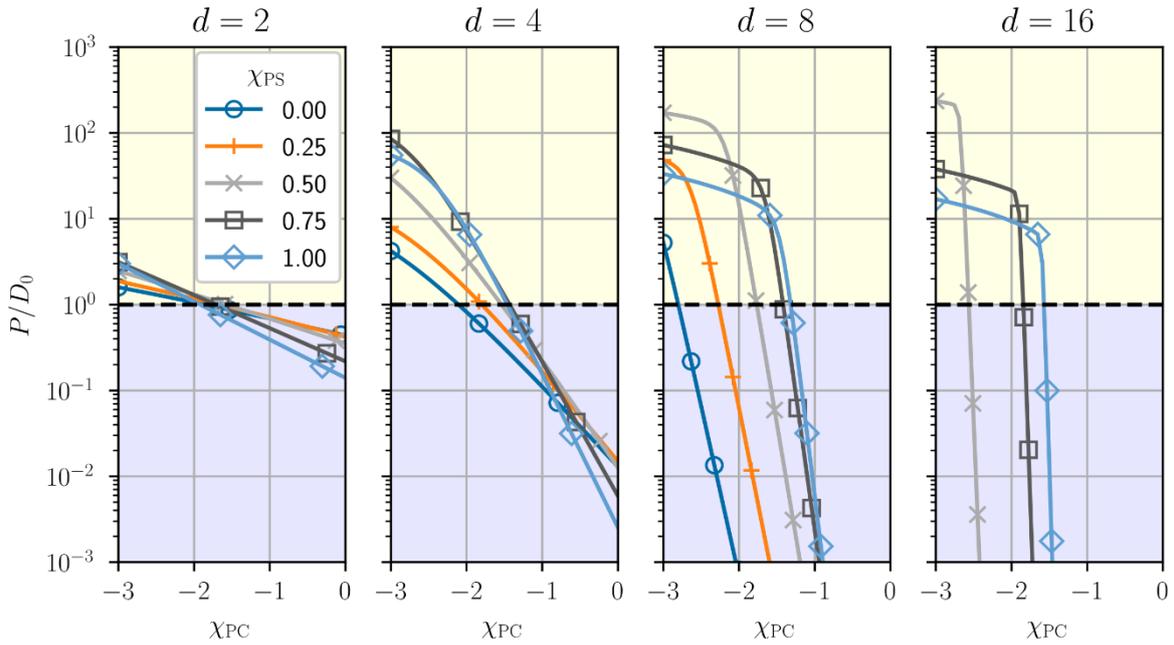

**Figure 7.** Effect of colloid-polymer interaction strength on brush permeability to colloids. Plots of the normalized permeability $P/D_0$ as a function of polymer-colloid interaction strength $\chi_{PC}$ for a set of solvent strengths ($\chi_{PS} = 0$, 0.25, 0.5, 0.75 and 1; symbols and colors are indicated in the legend in the left panel) and colloid sizes ($d = 2, 4, 8$ and $16$; as indicated atop the graphs). The regions of facilitated ($P/D_0 > 1$) and impeded ($P/D_0 < 1$) permeation are highlighted with light yellow and light blue backgrounds, respectively.

A major result emerging from our analysis is that polymer brushes are very effective in separating attractive colloids by their size and affinity to the polymer once the gating behavior in the near-critical regime is employed. **Figure 8** provides the threshold interaction strength defined by the critical condition that the brush is effectively transparent for the colloid, $P = D_0$, as a function of the solvent quality and colloid size. It is clear that, for a given polymer-colloid interaction strength $\chi_{PC}$, the size threshold (and *vice versa*) can be tuned over a wide range by adjusting the solvent strength. Thus, polymer brushes can serve as tuneable gates to select large enough colloids by their size, and by their affinity to the polymer.

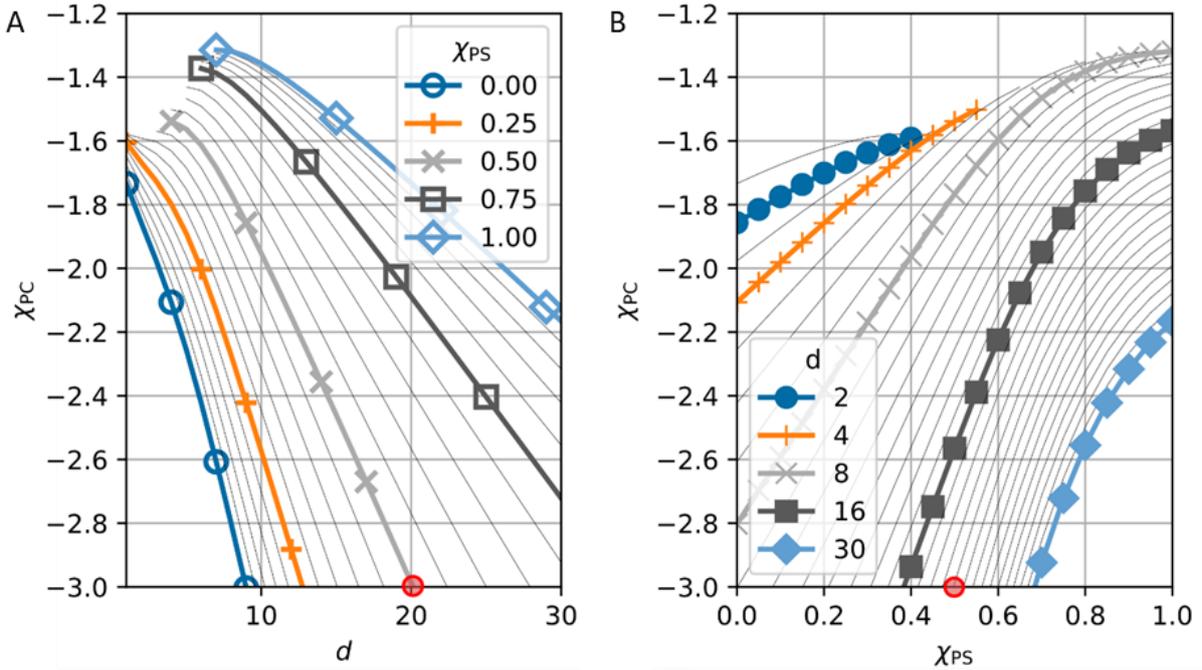

**Figure 8.** Mapping the parameter space for the critical condition. **(A)** Polymer-colloid interaction strength $\chi_{PC}$ as the function of the colloid size $d$ at the critical condition $P/D_0 = 1$ for a set of solvent strengths ($\chi_{PS} = 0$, 0.25, 0.5, 0.75 and 1.0, with symbols and colors as indicated in the legend; intermediate steps of 0.05 are shown by thin grey lines). **(B)** Polymer-colloid interaction strength $\chi_{PC}$ as the function of the solvent strength $\chi_{PS}$ at the critical condition $P/D_0 = 1$ for a set of colloid sizes ($d = 2, 4, 8, 16$ and 30, with symbols and colors as indicated; intermediate steps of 1 shown by thin grey lines). The red circles correspond to the same set of parameters ($\chi_{PS} = 0.5$, $\chi_{PC} = -3$, $d \approx 20$) as shown in **Figure 6**.

### 3.5. Effects of solvent strength and polymer-colloid affinity on colloid partitioning and connection to permeability

As derived in Section 2.6, the permeability is largely determined by the insertion free energy $\Delta F$, in particular for larger colloids. Another quantity that depends exclusively on the insertion free energy is the colloid partition coefficient, $c_{eq}/c_0 = e^{-\Delta F}$. To compare the effects of $\Delta F$ on colloid permeability and partitioning, respectively, we consider the mean partition coefficient $\langle c_{eq} \rangle/c_0$, as defined in Equation (11). **Figure 9** presents contour plots of the mean partition coefficient, arranged identically to **Figure 5** to facilitate side-by-side comparison.

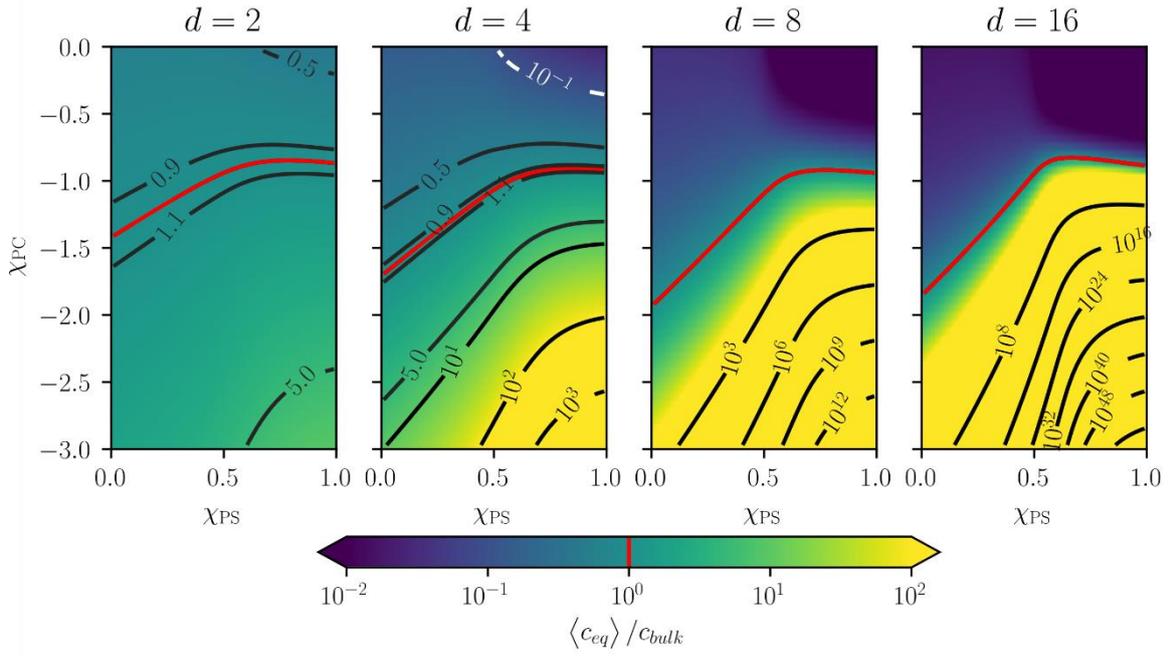

**Figure 9.** Effect of solvent strength and colloid interaction strength on colloid partitioning. Contour plots for the mean partition coefficient $\langle c_{eq}\rangle/c_0$ in the brush for a range of colloid sizes ($d = 2, 4, 8$ and $16$, as indicated atop the graphs) as a function of the interaction parameters $\chi_{PS}$ and $\chi_{PC}$. The solid red contours correspond to the equipartition condition $\langle c_{eq}\rangle/c_0 = 1$. Brush parameters: $N = 2000$, $\sigma = 0.08$.

Similar to the permeability, the partition coefficient shows extremely large variations with solvent strength and polymer-colloid affinity for the larger colloids ($d \geq 8$) and much smaller effects for smaller colloids ($d \leq 4$). The variations in the partition coefficient, however, are much more pronounced in the colloid accumulation regime ($\langle c_{eq}\rangle/c_0 \gg 1$) than in the colloid depletion regime ($\langle c_{eq}\rangle/c_0 \ll 1$). Whilst opposite to the permeability variations, these trends are also readily explained as due to differential effects of the brush interior and brush-solvent interface, as qualitatively described in Section 2.7: partitioning is dominated by the brush interior in the regime of colloid accumulation, with very strong effects, and by the brush periphery in the regime of colloid depletion, with rather weak effects.

We remind that the here-developed theory of colloid partitioning and brush permeability neglects colloid-colloid interactions. In practice, this implies that the colloid volume fraction should not exceed a value of the order of $10^{-1}$. For the larger colloids ($d \geq 8$), a part of the predicted data range will therefore be difficult to probe, as the colloid concentrations $c_0$ in the bulk solvent become either impractically small or too large to be able to probe the lower and upper right corners, respectively, of the parameter range in the contour plots in **Figure 9**.

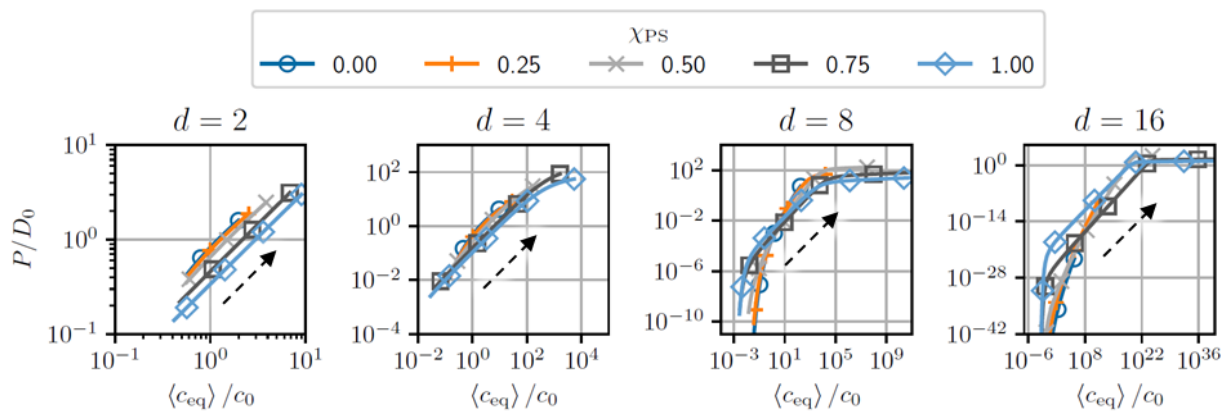

**Figure 10.** Linking colloid permeability to partitioning. Parametric plots of the normalized permeability $P/D_0$ as a function of the mean partition coefficient $\langle c_{eq}\rangle/c_0$ for a set of colloid sizes ($d = 2, 4, 8$ and $16$; as indicated atop the graphs) and solvent strengths ($\chi_{PS} = 0$, 0.25, 0.5, 0.75 and 1.0, with symbols and colors as indicated in the top legend). The dashed arrows indicate the direction of increasing polymer-colloid affinity (*i.e.*, decreasing $\chi_{PC}$) along each of the curves.

It was noted in Section 2.7 that if the brush density profile were perfectly uniform, and the brush-solvent interface effects completely absent leading to a uniform insertion free energy profile, the permeability and the partition coefficient would be linked by a simple proportionality relation, $P = D \langle c_{eq}\rangle/c_0$, which is well known and commonly used in membrane science. **Figure 10** presents parametric log-log plots of the normalized permeability, $P/D_0$, versus the average partition coefficient, $\langle c_{eq}\rangle/c_0$. In this plot, different curves correspond to different solvent qualities, $\chi_{PS}$, while each curve is parametrized by the polymer-colloid affinity, $\chi_{PC}$ (*i.e.*, each point on a given curve corresponds to a particular value of $\chi_{PC}$ which simultaneously defines the permeability and the average partition coefficient; the direction of increasing affinity, or decreasing $\chi_{PC}$, is indicated by dashed arrows). Naturally, both the permeability and the partition coefficient monotonically increase with increasing $\chi_{PC}$. Direct proportionality is revealed by the portions of the curves with the unit slope, most prominent in the two left panels showing the results for colloids of smaller size ($d = 2, 4$). The curves for larger colloids ($d = 8, 16$) also have slopes reasonably close to 1 in their central portions but even these portions cannot be recognized as representing the relation $P/D_0 = D/D_0 \langle c_{eq}\rangle/c_0$ because the observed proportionality coefficient between the reduced permeability and the average partition coefficient does not match the available estimates for $D/D_0$. Indeed, for $d = 8$ the central portion of the curve for $\chi_{PS} = 1$ with a slope close to 1 is well described by $P/D_0 \approx 10^{-3} \langle c_{eq}\rangle/c_0$ while the ratio $D/D_0$ is estimated as 0.03. For the largest colloid size, $d = 16$, the mismatch is even more drastic since the central portion of the curve for $\chi_{PS} = 1$ is described by $P/D_0 \approx 10^{-18} \langle c_{eq}\rangle/c_0$ while $D/D_0 \approx 0.01$. It is clear that the relation between the permeability and the colloid partitioning for large colloids is dramatically affected by the non-uniform insertion free energy profile. Indeed, Equations (20) and (11) show that while the brush permeability is most affected by the maximum of the insertion free energy profile due to the factor $e^{\Delta F(z)}$ in the integrand, the average partition coefficient contains the factor $e^{-\Delta F(z)}$ and hence is most affected by the minimum of the insertion free energy profile. The difference between the minimum and the maximum insertion free energies is large for colloids of larger size which explains the observed mismatch. The nearly horizontal or vertical tails observed in the two right panels of **Figure 10** are explicitly due to the brush-solvent interface effects as discussed in Section 2.7.

## 4. DISCUSSION AND CONCLUSIONS

We have developed a theory for the diffusive transport of colloids across polymer brushes, and explored the determinants of brush permeability across a wide range of solvent quality, colloid size and colloid affinity. In conclusion we would like to draw a qualitative picture that can be formulated in simple words. The focus of the present paper was on two experimentally measurable and closely related properties: the brush permeability under the assumption of transport with stationary flux, and the colloid partitioning between the solution and the brush under equilibrium conditions. The main findings are:

**1. A polymer brush can strongly affect colloid transport** compared to a solution layer of the same thickness. The most important factor is the average colloid insertion free energy $\Delta F$. One can identify two distinct regimes: the repulsive brush with $\Delta F > 0$ (impeded permeation), and the attractive brush with $\Delta F < 0$ (facilitated permeation). A secondary factor that always leads to slower transport is the reduced colloid mobility within the brush due to the polymer mesh effect. These qualitative results are illustrated in **Figure 11**.

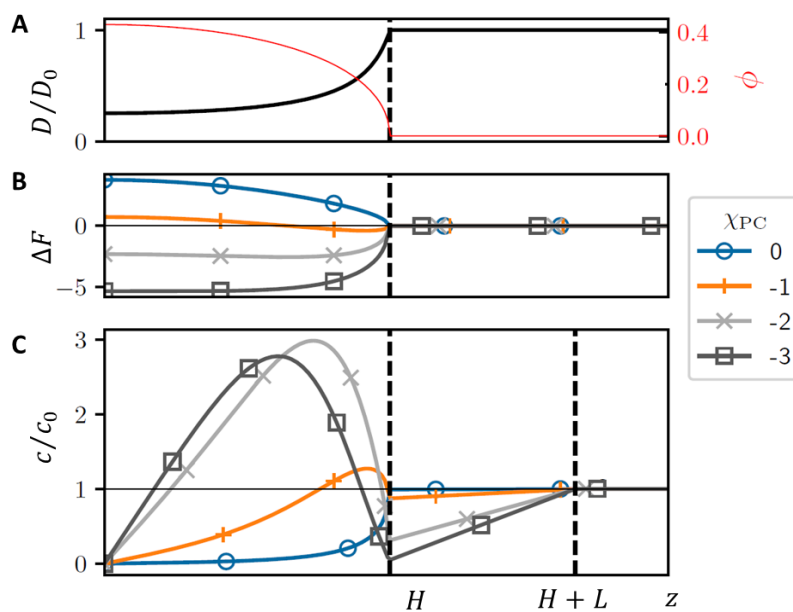

**Figure 11.** Normalized local diffusion coefficient, colloid insertion free energy and normalized colloid concentration profile for a representative polymer brush ($N = 2000$ and $\sigma = 0.08$) in $\theta$-solvent ($\chi_{PS} = 0.5$). **(A)** Local diffusion coefficient normalized by diffusion coefficient in the pure solvent (black solid line, left axis) and polymer density profile (red thin solid line, right axis). **(B)** Insertion free energy profiles for a colloid of size $d = 4$ and a set of polymer-colloid affinities ($\chi_{PC} = 0, -1, -2$ and $-3$; coded with colors and symbols as indicated next to the graphs). **(C)** Colloid concentration profiles within the brush and in the depletion layer normalized by the colloid concentration in the bulk solution for the same parameters as in **B**.

**2. Three key parameters control the insertion free energy:** the solvent's quality for the brush-forming polymer, the polymer-colloid affinity, and the colloid size. In order to have a strongly repulsive or a strongly attractive brush regime, the colloid size must be larger than the polymer Kuhn segment length by about an order of magnitude or more. Good solvent conditions typically produce a repulsive brush, mostly due to high osmotic pressure. The attractive brush regime is best achieved under $\theta$ or poor solvent conditions combined with pronounced adsorption of the polymer onto the surface of the colloid, with the affinity parameter well exceeding the critical value.

**3. Effects of the brush-solution interface.** Strongly repulsive brushes have an exponentially low permeability $P \sim e^{-\Delta F}$, and a rather low (non-exponential) average partition coefficient: the latter is dominated by the brush-solution interface where colloids are only partially immersed in the brush. Strongly attractive brushes have an exponentially large partition coefficient determined by the brush bulk properties, and a rather high (non-exponential) permeability which is dominated by the brush-solution interface.

**4. Effects of the boundary layer.** In an experimentally relevant situation, the colloid flux is affected not only by the brush properties but also by the thickness of the adjacent boundary solution layer, analogous to the situation when the net current is controlled by two resistors connected in series. The lower the brush resistance, the larger the impeding effect of the boundary layer. Hence, to realize enhanced transport through attractive brushes the thickness of the boundary layer ought to be small ($L \lesssim H$, which is typically in the sub-micrometer range). Biological cells provide a natural setting for such enhanced transport scenarios. Repulsive brushes are much less sensitive to the solution layer effects due their high intrinsic resistance.

**5. Permselectivity.** A polymer brush can be highly selective with respect to two colloid characteristics: size and affinity to the brush-forming polymer, but this requires that particles be large enough (diameter larger than 6 to 8 segment lengths). The most important parameter regime corresponds to near-critical condition when the free energy of insertion is close to zero, *i.e.*, at the boundary between the repulsive and the attractive brush regimes. The near-critical regime allows for high brush permeability and high selectivity at the same time, with the brush effectively blocking colloids larger than a certain threshold size or with an affinity below a threshold value. The gating thresholds can be tuned by changing the solvent quality. Polymer brushes can thus serve as tunable gates naturally suited for selecting large enough colloids by their size and affinity to the polymer.

**6. The graded brush density profile sensitively affects colloid permeability.** The limit of very thick homogeneous brushes where interfacial effects are negligible represents a well-studied case of transport through a macroscopic membrane where the membrane permeability and the equilibrium partition coefficient are proportional to each other: $P = D \langle c_{\text{eq}} \rangle / c_0$. This limit is approximately realized for smaller colloids ($d \leq 4$) under relatively poor solvent conditions. For larger colloids the non-homogeneous insertion free energy profile leads to strong deviations from the simple relation.

The novelty of the theoretical model developed here is that it enables exploration of a wide range of solvent qualities, which affects the conformation of the brush-forming chains, as well as colloid affinities and sizes. The model provides guidelines for the rational design of brushes tailored for specific applications in colloid sensing and separation, and for the mechanistic understanding of biological processes that involve colloid transport through polymer brushes and how these are optimized. Whilst we have used one example brush for illustration (Section 3.1), the main qualitative effects discussed above should be independent of the exact brush parameters (degree of polymerization and grafting density).

The model should also be a versatile starting point for the future development of extended models. A limitation of the current model, for example, is the treatment of colloids as non-interacting which restricts the validity of our results to relatively small colloid concentrations (volume fraction less than 0.1). Further, we have presented the data for spherical particles, although effects of the particle shape may bring some quantitative changes in the mapping of the near-critical regime due to different surface-to-volume ratios. Finally, the here-developed theoretical framework for planar brushes can in the future be built upon to analyze colloid transport through non-planar brushes, such as convex or concave spherical brushes (relevant to the outside or inside of polymersomes) and concave cylindrical brushes (relevant to nucleo-cytoplasmic transport).

**Acknowledgements**

This work was financially supported by the Russian Foundation for Basic Research (grant 21-53-10005), by the Royal Society (International Exchanges Award IEC/R2/202035 to R.P.R.), by the European Union's Horizon 2020 research and innovation program under the Marie Sklodowska-Curie grant agreement No 823883, and by the UK Biotechnology and Biological Sciences Research Council (grant BB/X00158X/1 to R.P.R.).

**Author contributions**

All authors have given approval to the final version of the manuscript. L.K., R.P.R. and O.V.B. conceived the study. All authors developed the theoretical model. M.Y.L. performed the computer simulations. All authors analyzed the data, and contributed to manuscript review and editing.

**References**

[1] W.-L. Chen, R. Cordero, H. Tran, C. K. Ober, *Macromolecules* **2017**, *50*, 4089.
[2] A. M. Bhayo, Y. Yang, X. He, *Prog Mater Sci* **2022**, *130*, 101000.
[3] B. W. Hoogenboom, L. E. Hough, E. A. Lemke, R. Y. H. Lim, P. R. Onck, A. Zilman, *Phys Rep* **2021**, *921*, 1.
[4] L. Mockl, *Front Cell Dev Biol* **2020**, *8*, 253.
[5] A. Varki, *Glycobiology* **2017**, *27*, 3.
[6] L. J. Stroh, T. Stehle, *Annu Rev Virol* **2014**, *1*, 285.
[7] T. Beddoe, A. W. Paton, J. Le Nours, J. Rossjohn, J. C. Paton, *Trends Biochem Sci* **2010**, *35*, 411.
[8] S. Weinbaum, J. M. Tarbell, E. R. Damiano, *Annu Rev Biomed Eng* **2007**, *9*, 121.
[9] J. A. Duncan, R. Foster, J. C. F. Kwok, *Br J Pharmacol* **2019**, *176*, 3611.
[10] S. Nourbakhsh, L. Yu, B. Y. Ha, *J Phys Chem B* **2021**, *125*, 8839.
[11] M. Badoux, M. Billing, H.-A. Klok, *Polym Sci* **2019**, *10*, 2925.
[12] L. Guo, Y. Wang, M. Steinhart, *Chem Soc Rev* **2021**, *50*, 6333.
[13] M. Kumar, M. Grzelakowski, J. Zilles, M. Clark, W. Meier, *Proc Natl Acad Sci U S A* **2007**, *104*, 20719.
[14] O. Onaca, R. Enea, D. W. Hughes, W. Meier, *Macromol Biosci* **2009**, *9*, 129.
[15] E. V. Konishcheva, U. E. Zhumaev, W. Meier, *Macromolecules* **2017**, *50*, 1512.
[16] N. Ben-Haim, P. Broz, S. Marsch, W. Meier, P. Hunziker, *Nano Lett* **2008**, *8*, 1368.
[17] P. Tanner, P. Baumann, R. Enea, O. Onaca, C. Palivan, W. Meier, *Acc Chem Res* **2011**, *44*, 1039.
[18] E. B. Zhulina, V. A. Priamitsyn, O. V. Borisov, *Polymer Science U.S.S.R.* **1989**, *31*, 205.
[19] C. M. Wijmans, J. M. H. M. Scheutjens, E. B. Zhulina, *Macromolecules* **1992**, *25*, 2657.
[20] M. Y. Laktionov, O. V. Shavykin, F. A. M. Leermakers, E. B. Zhulina, O. V. Borisov, *Phys Chem Chem Phys* **2022**, *24*, 8463.
[21] G. J. Fleer, M. A. Cohen Stuart, J. M. H. M. Scheutjens, T. Cosgrove, B. Vincent, "*Polymers at Interfaces*", Chapman & Hall, London, 1993.
[22] L. H. Cai, S. Panyukov, M. Rubinstein, *Macromolecules* **2011**, *44*, 7853.
[23] P.-G. de Gennes, "*Scaling Concepts in Polymer Physics*", Cornell University Press, Ithaca and London, 1979.
[24] H. Risken, "*The Fokker-Planck Equation. Methods of Solution and Applications.*", Springer, 1996.
[25] W. T. Hermens, M. Benes, R. P. Richter, H. Speijer, *Biotechn Appl Biochem* **2004**, *39*, 277.

**Table of Content entry**

We present a mean-field theoretical approach to analyze the binding and transport of colloids in planar polymer brushes. The position-dependent free energy of colloid insertion into a polymer brush is derived as a function of the solvent strength, the colloid size and the colloid-polymer affinity. We quantitate the brush permeability and show that brushes can serve as tunable gates for selecting colloids by their size and affinity for the polymer.

M. Y. Laktionov[1], E. B. Zhulina[1,2], L. Klushin[2,3], R. P. Richter[4,*], Oleg V. Borisov[1,2,5,*]

**Selective colloid transport across planar polymer brushes**

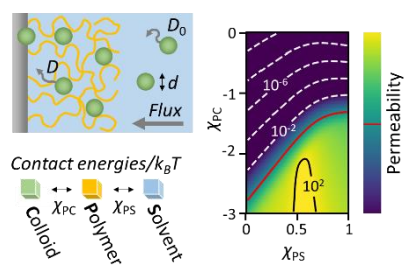

# Supporting Information

**Selective colloid transport across planar polymer brushes**

*Mikhail Y. Laktionov[1], Ekaterina B. Zhulina[1,2], Leonid Klushin[2,3], Ralf P. Richter[4], and Oleg V. Borisov[1,2,5],\**

**Supporting figures**

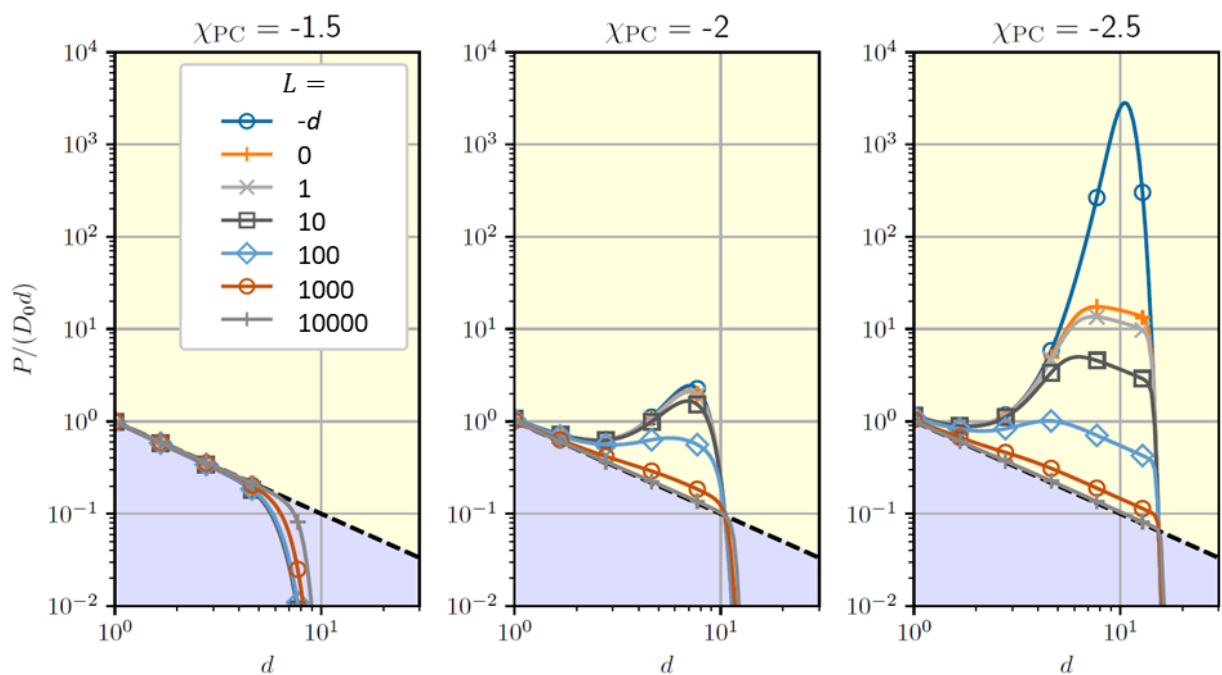

**Figure S1.** Effect of colloid size and depletion layer thickness on colloid permeability. Plots of the normalized permeability $P$ (as in **Figure 6**) as a function of colloid size $d$ for a selected solvent quality ($\theta$-solvent; $\chi_{PS} = 0.5$), three selected colloid interaction strengths ($\chi_{PC} = $ -1.5, -2, -2.5, as indicated atop the graphs), and a set of distances $L$ of the source of the colloids from the brush-solution interface grafting surface (as indicated with symbols and colors). For $\chi_{PC} = $ -1.5, the data for $L \leq 100$ virtually fall onto a single line. Brush parameters: $N = 2000$, $\sigma = 0.08$. The special condition $L = -d$ effectively neglects the impact of the brush-solution interface on the permeability (see **Figure 3**).